\documentstyle[12pt]{article}
\newcommand{\sbody}[2]{{\textstyle\frac{#1}{#2}}}
\begin{document}
\pagenumbering{arabic}
\begin{center}
\vfill
\large\bf{Supersymmetry in Spaces of}\\
\large\bf{Constant Curvature}
\end{center}
\vfill
\begin{center}
D.G.C. McKeon$^{(1)}$\\
Department of Applied \vspace{-.5cm}Mathematics\\
University of Western \vspace{-.5cm}Ontario\\
London\vspace{-.5cm}\\
CANADA $\;\;$ N6A 5B7\\
T.N. Sherry$^{(2)}$\\
Department of Mathematical \vspace{-.5cm}Physics\\
National University of \vspace{-.5cm}Ireland\\
Galway\vspace{-.5cm}\\
IRELAND
\end{center}
\vfill
$^{(1)}$email: DGMCKEO2@UWO.CA\\
$^{(2)}$email: TOM.SHERRY@NUIGALWAY.IE
\eject

\section{Abstract}

Supersymmetry is considered in spaces of constant curvature (spherical, de Sitter
and Anti-de Sitter spaces) of two, three and four dimensions.

\section{Introduction}

The expectation that supersymmetry will soon be shown to be a symmetry of nature motivates
us to analyze its properties in spaces of constant curvature. We view these $D$
dimensional spaces as being surfaces, in a flat embedding space
of dimension $D + 1$, that satisfy the (constraint) equation
$$g_{AB} x^A x^B = {\rm{const}} .\eqno(1)$$
For spherical space times $S_D$, $g_{AB} = {\rm{diag}}(+,+,+, \ldots
+)$, for de Sitter space $dS_D$, $g_{AB} = {\rm{diag}}(+,+,+, \ldots
+, -)$ and for anti de Sitter space, 
$AdS_D$, $g_{AB} = {\rm{diag}}(+ + + \ldots +, -, -)$. The symmetry 
(or isometry) transformations on this surface
are generated by operators $J^{AB}$ which satisfy the algebra
$$\left[ J^{AB}, J^{CD}\right] = g^{AC}J^{BD} - g^{BC}J^{AD} + g^{BD}
J^{AC} - g^{AD} J^{BC} .\eqno(2)$$
(We use anti-Hermitian generators.) Depending on the metric, this is the
algebra of the groups $S0(D+1)$, $S0(D, 1)$ and $S0(D-2, 2)$ for $S_D$,
$dS_D$ and $AdS_D$ respectively. We do not necessarily include Bosonic translation
generators $P_\mu$, in contrast to ref [1].

The isometry group of flat space-times is a 
semi-direct product of an $SO$ group, such as the above,
with the Abelian group of translations; for example 
in $3 + 1$ dimensions the Poincar\"{e} group is $SO(3,1) \times T_4$. 

The supersymmetry
generators $Q$ in flat space-time occur as ``square roots'' of the 
translation generators $P$ ($\left\lbrace Q, Q\right\rbrace \sim P$). The translation
generators $P^A$ commute with the supersymmetry generators $Q$ and the $SO$
generators $J^{AB}$.

In the case of the spaces of constant curvature translations are not isometries. The
supersymmetry generators $Q$ must then occur as ``square roots'' of the $J^{AB}$
($\left\lbrace Q, Q\right\rbrace \sim J^{AB}$); the $J^{AB}$ do not commute
with $Q$. Closure of the algebra requires in many cases extra Bosonic generators.
These do not necessarily commute with
everything and consequently are not ``central charges'' but rather
``internal symmetry generators''. This is in contrast to the central
charges considered in [2].

In section 3 we consider a number of different possible supersymmetry
algebras which extend the algebra of (2) for these spaces of constant
curvature for $D = 4, 3$ and $2$. In each case we have identified
extensions with the least possible number of Fermionic generators. (We
consider cases in which the simplest Fermionic generator is a Dirac
spinor, unlike ref. [1].)

In Sections 4, 5 and 6 we focus our attention on the simplest
supersymmetry algebras in the spaces $S_2$ and $AdS_2$.  In Section 4 we
examine the representations of the algebras.  For $S_2$ we classify the
states which carry an irreducible representation of the algebra and we
show that there is an upper bound on the angular momentum for these
states. For $AdS_2$ we generalise the previous treatment of the $N = 1$
supersymmetry algebra [8, 9, 16] to the $N = 2$ case. The $N = 2$
algebra in $AdS_2$ resembles closely the simplest supersymmetry algebra
in $S_2$. In section 5 we provide examples of supersymmetric models on
$S_2$ and $AdS_2$ containing interacting scalar and spinor fields. In
section 6 we provide a realization in appropriate superspaces of the
minimal supersymmetry algebra in $S_2$ and $AdS_2$ using the Bosonic
coordinates of the 3-dimensional embedding spaces. In the $AdS_2$ case
we define scalar superfields and we write down a number of distinct
supersymmetric actions in terms of them. In terms of the component
fields these actions correspond to realistic models on $AdS_2$.

Notation and Dirac matrix identities are given in an appendix.

\section{Supersymmetry Algebra}
\subsection{$S_4$}

As noted after eq. (A.36), in $5 + 0$ dimensions the simplest spinor is necessarily
Dirac[3]; we consequently employ a Dirac spinorial generator $Q_i$.
It is necessary to include a
scalar internal symmetry generator $Z$ in addition to $J^{AB}$ in order
to define a closed supersymmetric extension of the algebra of eq. (2) for
$S_4$. The two superalgebras whose non-vanishing (anti-)commutators are
$$\left[J^{AB}, Q_i\right] = -\Sigma_{ij}^{AB} Q_j\eqno(3a)$$
$$\left[Z, Q_i\right] = \mp Q_i\eqno(3b)$$
$$\left\lbrace Q_{i,} Q_j^\dagger \right\rbrace
= Z \delta_{ij} \pm \Sigma_{ij}^{AB} J^{AB}\eqno(3c)$$
satisfy the Jacobi identities. (Eq.
(A.34) is useful in proving this.)

Another pair of superalgebras associated with $S_4$ is given by
$$\begin{array}{rcl}
\left[J^{AB}, Q_i\right] = -\Sigma_{ij}^{AB} Q_j & &
\left[J^{AB}, Z^C\right] = \delta^{AC}Z^B - \delta^{BC} Z^A\end{array}\eqno(4a,b)$$
$$\begin{array}{rcl}
\left[Z^A, Q_i\right] = -\sbody12 \gamma_{ij}^A Q_j & &
\left[Z, Q_i\right] = -Q_i\end{array}\eqno(4c,d)$$
$$\left[Z^A, Z^B\right] = -J^{AB} \eqno(4e)$$
$$\left\lbrace Q_i , Q_j^\dagger \right\rbrace = \pm \left( \sbody32
Z\delta_{ij} - \gamma_{ij}^A Z^A + \Sigma_{ij}^{AB}
J^{AB}\right).\eqno(4f)$$
It too is consistent with the Jacobi identities. 
The superalgebras of equations (4) differ from those of equations (3)
through the inclusion of an $SO(5)$ vector bosonic generator $Z^A$ which
does not commute with $J^{AB}$ and a distinct anti-commutator 
$\left\lbrace Q, Q^\dagger\right\rbrace$.  $Z^A$ plays the role of a
translation operator as in ref. [1].
In (3) and (4), the
$J_{AB}$ are anti-Hermitian while $Z_A$ and $Z$ are Hermitian.

\subsection{$S_3$}

In $4 + 0$ dimensions also the simplest spinors are
Dirac [3]. Consequently one can obtain a superalgebra associated with
$S_3$ by ``dimensional reduction'' of the superalgebras of eq. (3) and
eq. (4). One identifies the generators $J^{\alpha 5}$ with an $S0(4)$
vector generator $Y^\alpha$. The superalgebra of (3) is then rewritten as
$$\begin{array}{rcl}
\left[ J^{\alpha\beta}, Q_i \right] = -\Sigma_{ij}^{\alpha\beta} Q_j
, & & \left[Y^\alpha , Q_i \right] = \sbody12 \left(\gamma^\alpha
\gamma_5\right)_{ij} Q_j\end{array}\eqno(5a,b)$$
$$\begin{array}{rcl}
\left[ J^{\alpha\beta}, Y^\gamma \right] = \left(\delta^{\alpha\gamma}
Y^\beta - \delta^{\beta\gamma} Y^\alpha\right),
& & \left[Y^\alpha , Y^\beta \right] = J^{\alpha\beta}
\end{array}\eqno(5c,d)$$
$$\begin{array}{rl}
\left[Z, Q_i \right] = \mp Q_i, & 
\left\lbrace Q_i, Q_j^\dagger \right\rbrace = Z\delta_{ij} \pm
\left(\Sigma_{ij}^{\alpha\beta} J^{\alpha\beta} - \sbody12 
 \left(\gamma^\alpha \gamma_5 \right)_{ij} Y^\alpha\right)
\end{array}\eqno(5e,f)$$
in addition to eq. (2).  Again, $Y$ is a translation operator, as in [1].

When ``dimensionally reducing'' the superalgebra of eq. (4), we again
identify $J^{\alpha 5}$ with $Y^\alpha$; in addition $Z^5$ becomes the
$S0(4)$ scalar $Y$. It is straightforward to effect the ``dimensional
reduction'' of the superalgebra of eq. (4); the approach employed in
obtaining the superalgebra of eq. (5) from that of eq. (3) is followed.

Of more interest, there are two superalgebras associated with $S_3$ which cannot
be obtained by ``dimensional reduction''. These superalgebras are the analogues of
equations (3). They require two internal symmetry generators $Z$ and
$Z_5$ to satisfy the Jacobi Identities. The superalgebras are given by
$$\left\lbrace Q_i, Q_j^\dagger \right\rbrace = 
\mp \Sigma_{ij}^{\alpha\beta} J^{\alpha\beta} + Z\delta_{ij} 
+ Z_5\left(\gamma_5\right)_{ij} \eqno(6a)$$
$$\left[ Z, Q_i \right] = \pm \sbody12 Q_i \;\;\;\; \left[ Z_5, Q_i
\right] = \pm \sbody12 (\gamma_5)_{ij} Q_j \eqno(6b,c)$$
$$\left[J^{\alpha\beta}, Q_i \right] = -\Sigma_{ij}^{\alpha\beta}
Q_j\eqno(6d)$$
in addition to eq. (2). This superalgebra can be further decomposed into two decoupled 
subalgebras, using the chiral decomposition of the generators $Q$ into $\frac{1}{2}
\left(1 \pm \gamma_5\right)Q$.

\subsection{$S_2$}

Superalgebras with one Fermionic generator associated with a two dimensional spherical
surface embedded in $3 + 0$ dimensions cannot be obtained by
dimensional reduction of the superalgebras associated with $S_3$, as in
$3 + 0$ dimensions irreducible spinors are two component Dirac spinors while in $4 + 0$ dimensions
they are four component Dirac spinors. Superalgebras, with one
Fermionic generator, similar in form to those of (3) and (6) can however be
written down on $S_3$. If $Q_i$ is a
two-component Dirac spinorial generator, then we find two superalgebras
consistent with the Jacobi identities, namely
$$\left\lbrace Q_i, Q_j^\dagger \right\rbrace = Z\delta_{ij} \mp 2
\tau_{ij}^a J^a\eqno(7a)$$
$$\left[J^a, Q_i \right] = -\sbody12 \tau_{ij}^a Q_j\eqno(7b)$$
$$\left[Z, Q_i\right] = \mp Q_i\eqno(7c)$$
$$\left[ J^a , J^b\right] = i\epsilon^{abc} J^c.\eqno(7d)$$

In addition there is in two dimensions a third superalgebra associated with
$S_2$. It can be shown that (with $\tilde{Q} = Q^T \tau_2$)
$$\begin{array}{rcl}
\left\lbrace Q_i, \tilde{Q}_j \right\rbrace = \tau_{ij}^a J^a & &
\left\lbrace Q_i, Q_j^\dagger \right\rbrace = \tau_{ij}^a
Z^a\end{array}\eqno(8a,b)$$
$$\begin{array}{rcl}
\left[J^a,  Q_i \, , \right] = -\sbody12 \tau_{ij}^a Q_j & &
\left[Z^a,\tilde{Q_i}\right] = \sbody12 Q_j^\dagger \tau_{ji}^a
\end{array}\eqno(8c,d)$$
$$\begin{array}{rcl}
\left[J^a,  J^b, \right] = i\epsilon^{abc}J^c & &
\left[Z^a, Z^b\right] = -i\epsilon^{abc} J^c
\end{array}\eqno(8e,f)$$
$$\left[J^a, Z^b \right] = i\epsilon^{abc} Z^c\eqno(8g)$$
is consistent with Jacobi identities. This algebra does not appear to
have an analogue on $S_3$ or $S_4$. 
If we define the symplectic Majorana spinors
$$Q_1 = \frac{Q+Q_c}{2} = -(Q_2)_c\nonumber$$
$$Q_2 = \frac{Q-Q_c}{2} = (Q_1)_c\nonumber$$
then it follows from (8a,b) that
$$\left\lbrace Q_1, \tilde{Q}_2 \right\rbrace = \left\lbrace Q_2 ,
\tilde{Q}_1 \right\rbrace = \sbody12 \tau^a J^a\nonumber$$
$$\left\lbrace Q_1, \tilde{Q}_1 \right\rbrace = -\left\lbrace Q_2 ,
\tilde{Q}_2 \right\rbrace = -\sbody12 \tau^a Z^a\nonumber$$
$Z^a$ is akin to the translation operator appearing in [1].
A similar decomposition of (7) (with the upper sign)
leads to 
$$\left\lbrace Q_i , \tilde{Q}_j \right\rbrace = 
- \tau^a_{ij} J^a  + \frac{1}{2} Z\epsilon_{ij}\nonumber$$

\subsection{$dS_4$}

By exploiting some of the properties of spinors in $4 + 1$ dimensions (as given in
the appendix) we are able to formulate a supersymmetric algebra
associated with $dS_4$. We introduce two 4-component Dirac spinorial
generators $Q_i^r\;(r = 1,2)$ related by the symplectic Majorana
condition (A.15),
$$Q_i^r = \epsilon^{rs} Q_{s\,i}^y.\eqno(9)$$

The generator $\tilde{Q}_i^r$ is defined to be
$$\tilde{Q}_i^r = \left(Q^{Tr} C\right)_i\eqno(10)$$
with $C = C^\dagger = C^{-1} = -C^T = -C^*$ given by eq. (A.38). It is
evident that $\left(\tilde{Q}^r Q^s\right)$ is a Lorentz scalar in $4 +
1$ dimensions.

A consistent superalgebra employing these spinors that involves
a translation operator $Z_A$ is given by
$$\left\lbrace Q_i^r, \tilde{Q}_j^s \right\rbrace = i\left[\delta^{rs}
\Sigma_{ij}^{AB} J_{AB} + \epsilon^{rs} \left(\gamma_{ij}^A Z_A +
\delta_{ij} Z \right)\right]\eqno(11a)$$
$$\left[J^{AB}, Q_i^r\right] = -\left(\Sigma^{AB}
Q^r\right)_i\eqno(11b)$$
$$\left[ Z^A, Q_i^r \right] = -\sbody12 \epsilon^{rs} \left(\gamma^A Q^s
\right)_i\eqno(11c)$$
$$\left[ Z, Q_i^r \right] = \sbody32 \epsilon^{rs} Q^s_i\eqno(11d)$$
$$\left[ Z^A, Z^B \right] = J^{AB}\eqno(11e)$$
$$\left[J^{AB}, Z^C\right] = g^{AC} Z^B - g^{BC} Z^A.\eqno(11f)$$
Showing that the Jacobi identities are satisfied involves using the
properties of Dirac matrices (A.28 - A.34). In addition, to analyze the
Jacobi identity associated with $\left(Q_i^r, Q_j^s, Q_k^t\right)$,
(A.41) is useful.  The algebra of eq. (11) is closed under Hermitian
conjugation with the symplectic Majorana condition (9), provided
$J_{AB}^\dagger = - J_{AB}$, $Z_A^\dagger = - Z_A$ and $Z^\dagger = -
Z$.  Of course, one could always use a single 4-component 
Dirac spinor in place of the symplectic
Majorana spinors (9). In that case, an analogue of either (3) and (4) could
be introduced in $dS_4$; similarly (11) has an analogue on $S_4$. A
mapping between these two algebras in the latter case between the
algebras of (11) and (4) is provided by
$\displaystyle{\frac{Q^1 + iQ^2}{\sqrt{2}} \rightarrow Q}$, $iZ^A
\rightarrow Z^A$, $Z \rightarrow \frac{3i}{2} Z$.

\subsection{$AdS_4$}

In the case of $AdS_4$, one can use spinorial
generators which are Majorana rather than Dirac, as the Majorana
condition can be consistently applied in $3 + 2$ dimensions.

If now $\tilde{Q} = Q^TC$ with $C$ defined by eq. (A.40), then we can
have
$$\left\lbrace Q_i , \tilde{Q}_j \right\rbrace = 2 \Sigma_{ij}^{AB}
J_{AB}\eqno(12a)$$
$$\left[ J^{AB}, Q_i \right] = - \left(\Sigma^{AB} Q
\right)_i\eqno(12b)$$
as a consistent supersymmetric generalization of the algebra associated with 
$AdS_4$.
Again, (A.41) is crucial for establishing that the Jacobi identities are
satisfied.

We note that (12a) is consistent in the sense that $\left(\Sigma^{AB}
C\right)^T = \Sigma^{AB} C$; vectorial and scalar internal symmetry
generators cannot consistently be introduced into (12a) as
$\left(\gamma^A C\right)^T = -\gamma^A C$, $C^T = -C$.

\subsection{$dS_3$}

Using the notation employed in the appendix  we note that the usual supersymmetry 
algebra for $3 + 1$ dimensional
Minkowski space, is given
by 
$$
\left\lbrace Q_i, \tilde{Q}_j \right\rbrace = 0 \;\;\;\;\;\;
\left\lbrace Q_i, Q_j^\dagger \right\rbrace =
2\sigma_{ij}^\lambda P_\lambda\nonumber$$
$$\left[P_\mu, \tilde{Q}_1 \right] = 0 \;\;\;\;\;\;
\left[ J_{\mu\nu}, Q_i\right] = -\left(\sigma_{\mu\nu} Q\right)_{i}\;\;\;\;\; 
\left[J_{\mu\nu}, Q_i^\dagger \right] = \left(Q^\dagger \overline{\sigma}_{\mu\nu}
\right)_i\eqno(13a)$$
$$\left[P_\mu, P_\nu \right] = 0 \;\;\;\;
 \left[ J_{\mu\nu}, P_\lambda\right] = g_{\mu\lambda} P_\mu - g_{\nu\lambda}
 P_\mu \;\;\;\;
\left[J_{\mu\nu}, J_{\lambda\sigma}\right] = g_{\mu\lambda} J_{\nu\sigma} - g_{\mu\sigma}
J_{\nu\lambda} + g_{\nu\sigma} J_{\mu\lambda} - g_{\nu\lambda} J_{\mu\sigma}.
\nonumber$$
Since no anticommutator of the form $\left\lbrace Q_i,
\tilde{Q}_j\right\rbrace$,  $\left\lbrace Q_i,
Q_j^\dagger\right\rbrace$ or $\left\lbrace Q_i^\dagger,
Q_j^\dagger\right\rbrace$ can consistenty be related to $J_{\mu\nu}$,
this algebra cannot be viewed as a superalgebra in $dS_3$. 
However, as in [1] if
we relax the condition $\left[P_\mu, P_\nu\right] = 0$ and identify $P_\mu$ with an internal
symmetry generator $Z_\mu$, then we can consistently relate
$\left\lbrace Q_i,
\tilde{Q}_j\right\rbrace$ and $\left\lbrace \tilde{Q}_i^\dagger,
Q_j^\dagger\right\rbrace$ to $J_{\mu\nu}$ so that the resulting
superalgebra
can be associated with $dS_3$. It is
$$
\left\lbrace Q_i, \tilde{Q}_j \right\rbrace = -2\sigma^{\mu\nu}_{ij} J_{\mu\nu}, 
\;\;\;
\left\lbrace \tilde{Q}^\dagger_i , Q^\dagger_j\right\rbrace = -2
\overline{\sigma}_{ij}^{\mu\nu} J_{\mu\nu}\;\;\;
\left\lbrace Q_i, Q_j^\dagger \right\rbrace =
2\sigma_{ij}^\lambda Z_\lambda\nonumber$$
$$\left[J_{\mu\nu ,}, Q_i \right] = -\left(\sigma_{\mu\nu} Q\right)_i\;\;\;\;\;\;
\left[ J_{\mu\nu ,} Q^\dagger_i \right] = \left(Q^\dagger \overline{\sigma}_{\mu\nu}\right)_i
\nonumber$$
$$
\left[Z_\mu, \tilde{Q}_i \right] =
\sbody12\left(Q^\dagger\overline{\sigma}_\mu\right)_i\;\;\;\;
 \left[ Z_\mu , Q_i^\dagger\right] = \sbody12
\left(\tilde{Q}\sigma_\mu\right)_i \eqno(13b)$$
$$
\left[Z_\mu, Z_\nu \right] = -J_{\mu\nu} \;\;\;
\left[J_{\mu\nu}, Z_\lambda \right] = g_{\mu\lambda} Z_\nu -
g_{\nu\lambda}
Z_\mu\;\;\;
\left[ J_{\mu\nu}, J_{\lambda\sigma}\right] = g_{\mu\lambda} J_{\nu\sigma}
- g_{\mu\sigma}J_{\nu\lambda} + g_{\nu\sigma} J_{\mu\lambda} -
g_{\nu\lambda}J_{\mu\sigma}
\nonumber$$
It can easily be shown that all Jacobi identities associated with the triples
$(Q, Q, Q)$, $(Q, Q, Q^\dagger)$, $(P, Q, Q)$, $(P, Q, Q^\dagger)$, 
$(P, P, Q)$, $(J, P, Q)$, $(P, P, J)$ and $(P, J, J)$ are satisfied
by the algebra of (13b).  This can be viewed as a supersymmetric extension of the algebra
considered in [4].

\subsection{$AdS_3$}

In $2 + 2$ dimensions, a spinorial generator $Q$ can be simultaneously Majorana
and Weyl. For the simplest supersymmetric extension of the isometry algebra of
$AdS_3$, we consider a spinorial generator in $2+ 2$ dimensions which is Majorana-Weyl. 
Using the notation of [5] with
$$\left( \sigma^{\mu\nu}\right)_{k\ell} = -\sbody14
\left[\sigma^\mu_{\;k\dot{m}} \overline{\sigma}^{\nu\dot{m}}_{\;\;\;\;\ell} -
\sigma^\nu_{\;k\dot{m}}
\overline{\sigma}^{\mu\dot{m}}_{\;\;\;\;\ell}\right]\eqno(14)$$
we find that
$$\left\lbrace Q_k ,Q^\ell \right\rbrace = \left(
\sigma^{\mu\nu}\right)_k^{\;\;\ell} J_{\mu\nu}\eqno(15a)$$
$$\left[J^{\mu\nu}, Q_k\right] = -
\left(\sigma^{\mu\nu}\right)_k^{\;\;\ell}Q_\ell\eqno(15b)$$
is a superalgebra which is consistent with the Jacobi identities.

\subsection{$AdS_2/dS_2$}

There is a degeneracy between two dimensional Anti-de Sitter space and
two dimensional de Sitter space. We begin by relating supersymmetry in $AdS_2$ space to
superconformed symmetry in $0 + 1$ dimensions [6].  The $N = 2$ superconformal algebra in
$0 + 1$ dimensions is given
by [7-9]
$$\left[ \pi , \delta \right] = \pi \;\;\;\;\; \left[ \kappa , \delta
\right] = -\kappa \;\;\;\;\; \left[\pi , \kappa \right] = 2\delta
\eqno(16a)$$
$$\left[ \delta , q_i \right] = -\sbody12 q_i \;\;\;\;\; \left[ \delta ,
s_i \right] = \sbody12 s_i \eqno(16b)$$
$$\left[ \pi , s_i \right] = i q_i \;\;\;\;\; \left[ \kappa , q_i
\right] = i s_i\eqno(16c)$$
$$\left\lbrace q_i , q_j \right\rbrace = \pm i\delta_{ij} \pi \;\;\;\;\;
\left\lbrace s_i, s_j \right\rbrace = \mp i \delta_{ij} \kappa
\eqno(16d)$$
$$\left\lbrace q_i , s_j \right\rbrace = \mp \delta_{ij} \delta -
\sbody{i}{2} \epsilon_{ij} \alpha \eqno(16e)$$
$$\left[ s_i , \alpha \right] = \mp i\epsilon_{ij} s_j\;\;\;\;\;
\left[q_i, \alpha \right] = \mp i\epsilon_{ij}q_j .\eqno(16f)$$
(This can be derived by projecting the $N = 2$ superconformal algebra in
$1 + 1$ dimensions  along the light cone [9].)  By setting $\alpha = q_2 = s_2 =
0$ in (16), a consistent $N = 1$ version of this superconformal algebra can be obtained. We note
that $i\delta$, $\pi$, $\kappa$, $\alpha$, $\sqrt{i} q_i$ and $\sqrt{-i}
s_i$ are Hermitian.

The generators $\pi$, $\delta$ and $\kappa$ (``Hamiltonian'',
``dilitation'' and ``special conformal'' generators respectively) have a
relationship with those of $AdS_2$ space given in [6]. We follow this
prescription, defining the symmetry generators $J_{ab} (= -J_{ba})$ in
$AdS_2$ space by
$$J_{12} = \sbody12(\kappa - \pi)\;\;\; J_{13} = \sbody12 (\kappa +
\pi)\;\;\; J_{32} = \delta .\eqno(17)$$
In addition, we define the two component spinor
$$Q = \left(\begin{array}{c}
q + is\\
q - is\end{array}\right).\eqno(18)$$
The $N = 1$ limit of supersymmetry algebra of (16) implies that 
$$\left[ J_{ab}, J_{cd} \right] = g_{ac} J_{bd} - g_{bc} J_{ad} + g_{bd}
J_{ac} - g_{ad} J_{bc}\nonumber$$
$$\left\lbrace Q , \tilde{Q} \right\rbrace = 2 \Sigma^{ab}
J_{ab}\eqno(19)$$
$$\left[ J_{ab}, Q \right] = -\Sigma_{ab} Q.\nonumber$$

If now we look at the full $N = 2$ superalgebra of (16) with
$$Q_i = \left( \begin{array}{c}
q_i + is_i\\
q_i - is_i\end{array}\right)\eqno(20)$$
then (16) implies that
$$\left\lbrace Q_i , \tilde{Q}_j \right\rbrace = -i\epsilon_{ij}
\alpha \pm 2\delta_{ij} \Sigma^{ab} J_{ab}\nonumber$$
$$\left[ J_{ab}, Q_i \right] = -\Sigma_{ab} Q_i\eqno(21)$$
$$\left[ \alpha , Q_i \right] = \pm i\epsilon_{ij} Q_j .\nonumber$$

From (18) and (20) it is apparent that the spinor generators in the
algebras (19) and (21) each possess two degrees of freedom, and so are
not true Dirac spinors.  In the algebras (19) and (21) $Q$ and $Q_i$ can
be interpreted as being Majorana spinors.
The two Majorana spinors in (21) can be combined to form a
Dirac spinor
$$Q = Q_1 + iQ_2 \; ;\eqno(22)$$
in this case (21) becomes
$$\left\lbrace Q, \tilde{Q} \right\rbrace = 0\nonumber$$
$$\left\lbrace Q , \overline{Q} \right\rbrace = \mp 4 \Sigma^{ab}
J_{ab} + 2\alpha\eqno(23)$$
$$\left[ J_{ab}, Q\right] = -\Sigma_{ab} Q\nonumber$$
$$\left[ \alpha , Q\right] = \pm Q.\nonumber$$

Using a Dirac spinor generator $Q$, we can formulate another consistent
superalgebra that is an extension of the $AdS_2$
algebra.
$$\left\lbrace Q, \tilde{Q}\right\rbrace = \Sigma^{ab} J_{ab}\;\;\;\;\;
\left\lbrace Q, \overline{Q}\right\rbrace = \Sigma^{ab}
Z_{ab}\nonumber$$
$$\left[J^{ab} , Q\right] = -\Sigma^{ab} Q \;\;\;\;\; \left[ Z^{ab},
\tilde{Q} \right] = \overline{Q} \Sigma^{ab}\eqno(24)$$
$$\left[ J^{ab}, J^{cd} \right] = g^{ac} J^{bd} - g^{bc} J^{ad} + g^{bd}
J^{ac} - g^{ad}J^{bc}\nonumber$$
$$\left[ Z^{ab}, Z^{cd} \right] = g^{ac} J^{bd} - g^{bc} J^{ad} + g^{bd}
J^{ac} - g^{ad}J^{bc}\nonumber$$
$$\left[ J^{ab}, Z^{cd} \right] = g^{ac} Z^{bd} - g^{bc} Z^{ad} + g^{bd}
Z^{ac} - g^{ad}Z^{bc}.\nonumber$$
This superalgebra is analogous to the $S_2$ superalgebra of eq. (8). It
 can be shown to satisfy the
Jacobi identities. On $AdS_2$, $Q$ can be decomposed into two
Majorana spinors
$$Q_{1,2} = \frac{Q \pm Q_c}{2}. \eqno(25)$$
The algebra of (24), in terms of $Q_{1,2}$, becomes
$$\left\lbrace Q_1 , \tilde{Q}_2 \right\rbrace = 0\nonumber$$
$$\left\lbrace Q_{1,2} , \tilde{Q}_{1,2} \right\rbrace = -\sbody12 \Sigma^{ab}
\left(Z_{ab} \mp J_{ab}\right)\eqno(26)$$
$$\left[ J^{ab}, Q_{1,2} \right] = -\Sigma^{ab}Q_{1,2}\nonumber$$
$$\left[Z^{ab}, Q_{1,2} \right] = \pm \Sigma^{ab} Q_{1,2}\nonumber$$
$$\left[ K_{\pm}^{ab}, K_{\pm}^{cd} \right] = g^{ac} K_{\pm}^{bd} - g^{bc} K_{\pm}^{ad}
+ g^{bd} K_{\pm}^{ac} - g^{ad} K_{\pm}^{bc}\nonumber$$
$$\left[ K_{\pm}^{ab} , K_{\mp}^{cd} \right] = 0\nonumber$$
where $K_{\pm}^{ab} = \sbody12 \left(J^{ab} \pm Z^{ab} \right)$.
It is evident from (26) that $Q_1$ and $Q_2$ both belong to a subalgebra with the
structure of eq. (19).

\section{Representations in Two Dimensions}
\subsection{$S_2$}

The analysis of the representations of the superalgebra of eq. (7) which is associated
with $S_2$ closely follows the discussion of the $sp\ell(2,1)$ superalgebra in [10].
(The $osp(2,1)$ superalgebra considered in [10] is not self-adjoint.) We
first note that there are two Casimirs associated with (7) (with the
upper sign)
$$\not\!\!C_2 = \vec{J}^{\,2} - \sbody14 Z^2 - \sbody12 Z + \sbody12
Q^\dagger Q\nonumber$$
$$ = \vec{J}^{\,2} - \sbody14 Z^2 - \sbody12 \left[ Q_i ,
Q_i^\dagger\right]\eqno(27)$$
and
$$\not\!\!C_3 = \sbody12 (Z + 1)\left(\not\!\!C_2 + \sbody12
Q^\dagger Q\right) +  \sbody14 Q^\dagger \tau \cdot JQ .\eqno(28)$$
The subalgebra of (7d) has the usual Casimir $\vec{J}^{\,2}$.

If $|I> = |j, m, \zeta >$ with
$$\vec{J}^{\,2} |I> = j(j + 1) |I> \eqno(29a)$$
$$J_3 |I> = m|I> \eqno(29b)$$
$$Z |I> = \zeta |I> \eqno(29c)$$
subject to the requirement
$$Q_i |I> = 0\;\;\;\;(i = 1, 2),\eqno(30)$$
we then define
$$Q_i^\dagger |I> = |i> \;\;\;\;(i = 1, 2)\eqno(31a)$$
$$Q_1^\dagger Q_2^\dagger |I> = |F>.\eqno(31b)$$

From (7) it follows that
$$J_3 |1> = \left(m - \sbody12 \right) |1> \eqno(32a)$$
$$J_3 |2> = \left(m + \sbody12 \right) |2> \eqno(32b)$$
$$J_3 |F> = m |F> ; \eqno(32c)$$
furthermore
$$Z |i> = \left(\zeta \pm 1\right) |i> \eqno(33a)$$
$$Z |F> = \left(\zeta \pm 2\right) |F>. \eqno(33b)$$
(The two signs in (33) correspond to the two algebras in (7).)

It is possible to show that
$$\left[J^2 , Q_1^\dagger Q_2^\dagger \right] = 0 \eqno(34)$$
and hence
$$J^2 |F> = j(j - 1)|F> \,; \eqno(35)$$
however, $|i>$ is a linear combination of states which are
eigenfunctions of $J^2$ corresponding to eigenvalues $(j + \sbody12 )(j
+ \sbody32)$ and $(j - \sbody12 )(j + \sbody12 )$.

(The operators appearing in (7) are related to those appearing in the
discussion of $sp\ell(2,1)$ in [11] by $H = J_3$, $E^\pm = J_1 \pm
iJ_2$, $\sqrt{2} F^+ = Q_1^\dagger$, $\sqrt{2} F^- = -
Q_2^\dagger$,
$\sqrt{2}\, \overline{F}^+ = -Q_2$, $\sqrt{2}\, \overline{F}^- = Q_1$
when the upper sign in (7) is used.)

Norms of states can be computed; we find
$$<1|1> = <I| \left\lbrace Q_1 , Q_1^\dagger \right\rbrace |I> =
(\zeta \mp 2m) <I|I>\eqno(36a)$$
$$<2|2> = (\zeta \pm 2m) <I|I> \eqno(36b)$$
$$<F|F> = (\zeta \mp 2j) (\zeta \pm 2j \pm 2) <I|I>\eqno(36c)$$
as
$$\left(J_1 \pm iJ_2 \right) |j,m> = \left[(j \mp m) (j \pm m +
1)\right]^{1/2} |j, m \pm 1> .\eqno(37)$$
For the norm of these states to be positive definite, we must have 
$$2j < \zeta \eqno(38)$$
so that $\zeta$ forms an upper bound on $j$.  This is similar to the
way in which the central charge forms an upper bound on the magnitude of
the momentum in the supersymmetric extension of the Poincar\'{e} group
in $4 + 0$ dimensions and $5 + 0$ dimensions [3,12].  By way of contrast, in the supersymmetric
extension of the Poincar\'{e} group in $3 + 1$ dimensions or $4 + 1$ dimensions, the central
charge forms a lower bound on the magnitude of the momentum [2,13-15,12].

\subsection{$AdS_2$}

Representations of the $N = 2$ superalgebra of (22) and (19) (and
consequently of (16)) can be worked out applying techniques used in [8]
for the $N = 1$ superalgebra.  The $N = 2$ superalgebra is also
considered in [9] and [16]; in the latter there is no analogue of the
operator $\alpha$ which is essential for the Jacobi identities.

Working directly with the operators of (16) (using the upper sign), the
subalgebra of (16a) has a Casimir
$$\not!\!C_0 = \delta^2 - \sbody12 \left\lbrace \pi , \kappa
\right\rbrace\eqno(39)$$
while the full superalgebra of (16) has the Casimir
$$\not\!\!C =\, \not\!\!C_0 + A - \sbody14 \alpha^2 \eqno(40)$$
where
$$A = A_1 + A_2 \eqno(41)$$
with
$$A_i = -\sbody12 \left[ q_i , s_i \right]\;\;\; (i = 1, 2) .\eqno(42)$$
Since
$$4A_1^2 - 2A_1 =\, \not\!\!C_0 = 4A_2^2 - 2A_2\eqno(43)$$
we see that
$$\not\!\!C = 2\left( A_1^2 + A_2^2\right) - \sbody14 \alpha^2
.\eqno(44)$$
A general discussion of the $S0(2,1)$ group of (16a) appears in [17].
The $S0(2)$ subgroup of $S0(2,1)$ has a Casimir
$$R = \sbody12 (\kappa - \pi ).\eqno(45)$$
We can now classify states $|\psi >$ by eigenvalues of $R$,
$\not\!\!C_0$, $\not\!\!C$ and $\alpha$ (taken to be $\rho$, $\gamma_0$,
$\gamma$ and $a$ respectively).  From (43) we see that
$$A_i|\psi > = A_i |\rho, \gamma_0, \gamma, a> = \frac{1 +
\epsilon_i\sqrt{1 + 4\gamma_0}}{4} \, |\rho , \gamma_0, \gamma , a>
\eqno(46)$$
where $\epsilon_i = \pm 1$; thus by (41)
$$\gamma = \gamma_0 = \sbody14 \left( 2 + \left(\epsilon_1 +
\epsilon_2\right) \sqrt{1 + 4\gamma_0} - a^2\right).\eqno(47)$$
Ladder operators for $S0(2,1)$ are given by
$$B_\pm = \sbody12 (\kappa + \pi ) \mp \delta = B_\mp^\dagger
;\eqno(48)$$
in addition, there are Fermionic ladder operators
$$F_\pm^i = q^i \pm is_i .\eqno(49)$$
These operators are related by
$$\not\!\!C_0 = R^2 - \sbody12 \left\lbrace B_- , B_+ \right\rbrace
\eqno(50a)$$
$$\left[ R, B_\pm \right] = \pm B_{\pm}\eqno(50b)$$
$$\left[R, F^i_\pm \right] = \pm \sbody12 F_\pm^i \eqno(50c)$$
$$\left\lbrace F_+^i, F_-^j \right\rbrace = -2i\delta_{ij} R -
\epsilon_{ij} \alpha\eqno(50d)$$
$$\left\lbrace F_\pm^i, F_\pm^j \right\rbrace = \sbody{i}{2}
\delta_{ij} B_\pm\eqno(50e)$$
$$\left[F_\pm^i, \alpha\right] = -i \epsilon_{ij} F_\pm^j\eqno(50f)$$
$$A_i = \sbody{i}{4} \left[F_-^i , F_+^i \right]\eqno(50g)$$
$$B_\pm B_\mp = R^2 \mp R - \not\!\!C_0\eqno(50h)$$
$$F_\pm^i F_\mp^i = -2iR \mp 2iA\eqno(50i)$$
$$\hspace{3.5cm} = -2iR \mp 2i\left(\not\!\!C - \not\!\!C_0 + \sbody14
\alpha^2\right)\nonumber$$
$$\left[ B_\pm ,F_\pm^i \right] = 0\eqno(50j)$$
$$\left[B_\pm , F_\mp^i \right] = \pm F_\pm^i .\eqno(50k)$$
If now we define
$$|\psi \pm> = B_\pm |\psi > \eqno(51a)$$
$$|\psi i \pm> = F_\pm^i |\psi > \eqno(51b)$$
then it follows that
$$\not\!\!C |\psi \pm > = \gamma |\psi \pm >\eqno(52a)$$
$$\not\!\!C_0 |\psi \pm > = \gamma_0 |\psi \pm >\eqno(52b)$$
$$R |\psi \pm > = (\rho \pm 1) |\psi \pm >\eqno(52c)$$
$$\alpha |\psi \pm > = a  |\psi \pm >\eqno(52d)$$
as well as
$$\not\!\!C |\psi i \pm > = \gamma |\psi i\pm >\eqno(53a)$$
$$R |\psi i \pm > = (\rho \pm \sbody12) |\psi i \pm >\eqno(53b)$$
$$\alpha |\psi i \pm > = \left(+i\epsilon_{ij} + a\delta_{ij}\right) |\psi j \pm >\eqno(53c)$$
$$\left( A + \not\!\!C_0\right) |\psi  i\pm > = \left( \not\!\!C_0 + \sbody14 \alpha^2\right)
|\psi i \pm > \eqno(53d)$$
$$\hspace{4cm}= \left[\left( + \sbody14 + \gamma + \sbody14 a^2\right) \delta_{ij} +
\sbody{i}{2} a\epsilon_{ij}\right] |\psi j \pm> .\nonumber$$
(The $S0(2,1)$ invariant $A + \not\!\!C_0$ is easier to work with than
$\not\!\!C_0$ itself as $\left[\not\!\!C_0 , F_\pm^i\right]$ is 
non-trivial while $\left[A + \not\!\!C_0,\; F_\pm^i\right]$ is easily
evaluated.)  One can now diagonalzie the $2 \times 2$ matrices on the
right side of (53c) and (53d).

Using (50h) and (50i), it follows that
$$<\psi\pm | \psi\pm > = \left(\rho^2 \pm \rho - \gamma_0\right) < \psi |
\psi > \eqno(54a)$$
and
$$<\psi i \pm | \psi i \pm > = \left[-2\rho \mp 2\left( \gamma - \gamma_0
+ \sbody14 \alpha^2 \right)\right] <\psi | \psi >.\eqno(54b)$$
The forms for $\gamma_0$ and $\rho$ are [17]
$$\gamma_0 = \Phi (\Phi + 1)\eqno(55a)$$
$$\rho = E_0 + n\;\;\;\; \left(-\sbody12 \leq E_0 < \sbody12 ,
n\;\;{\rm{integer}}\right)\eqno(55b)$$
with the permitted values of $\Phi$ and $E_0$ falling into four distinct
classes.  (In [8] the allowed values of $\Phi$ and $E_0$ are restricted
by an additional $0(3)$ symmetry present in the physical system being
considered.)

\section{Models in Two Dimensions}

It is possible to write down component field models in two dimensions
that are invariant under the supersymmetry transformations associated
with the superalgebras of (7), (19) and (21).  They bear a certain
resemblance to the hyperspherical models considered in [18-30].

\subsection{$S_2$}

It is possible to present a model invariant under transformations
associated with the algebra of eq. (7) (with the upper sign). We
consider the action
$$S = \int \frac{dA}{R^2} \Bigg[ \Big( \sbody12 \Psi^\dagger
\left(\vec{\tau}\cdot \vec{L} + x\right) \Psi - \Phi^*\left(L^2 + x(1-
x)\right)\Phi\nonumber$$
$$\left. - \sbody14 F^*F\right) + \lambda_N\Big(2(1-
2x)\Phi^*\Phi\nonumber$$
$$- (F^*\Phi + F\Phi^*) - \Psi^\dagger\Psi \Big)^N \Bigg]
.\eqno(56)$$
In (56), $\Phi$ and $F$ are complex scalars, and $\Psi$ is a Dirac
spinor, defined on the surface of a sphere of radius $R$ in three
dimensions. The angular momentum operator is $\vec{L} = -i \vec{r}
\times \vec{\nabla}$ and $x$ and $\lambda_N$ are arbitrary real
parameters. 

By using the identities of eqs. (A.1 - A.4), one can verify that for
arbitrary $N$, (56) is invariant under both the supersymmetry
transformation
$$\delta\Phi = \xi^\dagger\Psi\nonumber$$
$$\delta\Psi = 2\left(\vec{\tau} \cdot \vec{L} + 1 - x\right)\Phi \xi -
F \xi\eqno(57)$$
$$\delta F = -2 \xi^\dagger \left(\vec{\tau} \cdot \vec{L} +
x\right)\Psi\nonumber$$
and the special transformations
$$\delta\Psi = \lambda i\left(1 + 2 \vec{\tau} \cdot
\vec{L}\right)\Psi\nonumber$$
$$\delta\Phi = \lambda i\left(2 (1-x) \Phi - F\right)\eqno(58)$$
$$\delta F = \lambda i\left[ - 4\left(L^2 + x(1-x)\right)\Phi +
2xF\right]\nonumber$$
where $\xi$ is a constant Grassmann spinor and $\lambda$ is a
constant.  These transformations are generated by $\exp \left[ \xi^\dagger
R - R^\dagger \xi + i \lambda Z + i \vec{\omega} \cdot \vec{J}\right]$ using
the commutators
$$\left[\xi^\dagger R, \Phi \right] = \xi^\dagger \Psi\;\; ,
\;\;\;\;\;\left[\xi^\dagger R, \Psi \right] = \left[2(\vec{\tau} \cdot
\vec{L} + 1 - x) \Phi - F\right]\xi\nonumber$$
$$\left[\xi^\dagger R, F\right] = -2\xi^\dagger (\vec{\tau} \cdot
\vec{L} + x)\Psi\eqno(59)$$
$$\left[\lambda Z, \Phi \right] = -\lambda\left[2(1-x)\Phi - F\right]
,\;\;\;\;\; \left[\lambda Z, \Psi\right] = -\lambda\left[1 + 2\vec{\tau} \cdot
\vec{L}\right]\Psi\nonumber$$
$$\left[\lambda Z, F \right] = -\lambda\left[-4\left(\vec{L}^2 + x(1-x)\right)
\Phi + 2xF\right]\nonumber$$
$$\left[\vec{\omega}\cdot \vec{J} , \Phi \right] = -\vec{\omega} \cdot
\vec{L} \Phi ,\;\;\;\; \left[\vec{\omega} \cdot \vec{J}, \Psi\right] = -
\vec{\omega} \cdot \left( \vec{L} + \sbody12 \vec{\tau}
\right)\Psi\nonumber$$
$$\left[\vec{\omega} \cdot \vec{J} , F\right] = -\vec{\omega} \cdot
\vec{L} F.\nonumber$$
(All other commutators vanish.) Jacobi identities involving (7) (upper
sign) and (59) are satisfied.

Other models on $S_2$ also exist which possess a Fermionic symmetry. For
example let us consider
$$S = \int \frac{dA}{R^2} \left[i \Psi^\dagger \vec{\tau} \cdot \vec{r}
\left(\vec{\tau} \cdot \vec{L} + 1\right) \Psi + \Phi^* \vec{L}^2
\Phi\right].\eqno(60)$$
This differs in form from the action considered in [21] by the factor of
$\dot{i}$ in the first term (needed to ensure Hermiticity as $\vec{\tau}
\cdot \vec{r} \left(\vec{\tau} \cdot \vec{L} + 1\right) = - \left(\vec{\tau} \cdot \vec{L} + 1\right)
\vec{\tau} \cdot \vec{r}$). The action of eq. (60) possess the
superinvariance
$$\delta\Phi = R^2 \xi^\dagger \Psi\;\;\;\;\;\; \delta\Psi = -
i\vec{\tau} \cdot \vec{r} \vec{\tau} \cdot \vec{L} \Phi \xi\eqno(61)$$
where $\xi$ is again a Grassmann spinor. This symmetry however does not
appear to be consistent with the superalgebra of (7).

As a second model, let us take
$$S = \int \frac{dA}{R^2} \left[\sbody12 \Psi^\dagger \left(\vec{\tau} \cdot \vec{L}
 + x\right) \Psi + \Phi^* \left(\vec{L}^2 + x(3-x)-2\right)\Phi\right].\eqno(62)$$
This is invariant under
$$\delta\Phi = \xi^\dagger \vec{\tau} \cdot \vec{r} \Psi\;\;\;\;\;\;
\delta\Psi = -2\left(\vec{\tau} \cdot \vec{L} + 3 - x\right)\vec{\tau}
\cdot \vec{r} \Phi \xi\eqno(63)$$
as $\left[\vec{\tau} \cdot \vec{r} , L^2\right] = -2 \vec{\tau} \cdot
\vec{r} \left( \vec{\tau} \cdot \vec{L} + 1\right)$, but once again this
symmetry does not appear to be consistent with (7).   We note that
$\vec{\tau} \cdot \vec{r}$ has many of the properties of $\gamma_5$
in Euclidean space [31,32].

\subsection{$AdS_2$}

Models associated with the superalgebras of (19) and (21) can also be
devised.  The model
$$S = \int \frac{dA}{R^2} \left[ \tilde{\Psi} \left(\Sigma^{ab} L_{ab} +
\chi\right)\Psi + \Phi\left(\sbody12 L^{ab}L_{ab}\right.\right.\eqno(64)$$
$$\left.\left. + \chi (1 +\chi) \Phi\right) - F^2 +
\lambda_N\left((1+2\chi)\Phi^2 + 2\Phi F +
\tilde{\Psi}\Psi\right)^N\right]\nonumber$$
is invariant under
$$\delta\Psi = \left[\left( \Sigma^{ab}L_{ab} - (1 + \chi)\right)\Phi -
F\right]\xi\nonumber$$
$$\delta \Phi = \tilde{\xi}\Psi\;\; , \delta F = -
\tilde{\xi}\left(\Sigma^{ab}L_{ab} + \chi\right)\Psi\eqno(65)$$
if $L_{ab} = -x_a\partial_b + x_b\partial_a$. (Note that
$\left(\Sigma^{ab} L_{ab}\right)^2 = -\sbody12 L^{ab}L_{ab} +
\Sigma^{ab}L_{ab}$.) The superalgebra of (19) is consistent with the
transformations of (65).

In conjunction with the superalgebra of (21), (an $N = 2$ supersymmetry
algebra), one has an invariant model whose action is
$$S = \int \frac{dA}{R^2} \left[ \overline{\Psi} \left(\Sigma^{ab} L_{ab} +
\chi\right)\Psi + \Phi^*\left(\sbody12 L^{ab}L_{ab}\right.\right.\nonumber$$
$$\left.\left. + \chi (1 +\chi) \right)\Phi - F^*F\right]\eqno(66)$$
with now
$$\delta \Psi = \left[\left(\Sigma^{ab} L_{ab} - (1 + \chi)\right)\Phi -
F\right]\xi\eqno(67)$$
$$\delta\Phi = \overline{\xi}\Psi \;\; , \delta F = -\overline{\xi}
\left(\Sigma^{ab}L_{ab} + \chi\right)\Psi.\nonumber$$
In (64) $\Psi$ is a Majorana spinor and $\Phi$ and $F$ are real scalars;
in (64) $\Psi$ is a Dirac spinor
and $\Phi$ and $F$ are complex spinors.

\section{Superspace for Two Dimensions}

We can realize the two dimensional
superalgebras of eqs. (7) and (19) in superspace. Superspace models
invariant under (19) can be formulated.
\subsection{$S_2$}

In conjunction with the two superalgebras of (7), we consider a
superspace with coordinates $x^a$, $\theta_i$ and $\theta_i^\dagger$
where $\theta_i$ is a two component Dirac Grassmann spinor. In addition,
we employ an auxiliary constant $\beta$. The two superalgebras of (7)
have a representation [33]
$$Q = \left(\vec{\tau} \cdot \vec{x} + \beta\right)
\frac{\partial}{\partial\theta^\dagger} \pm
\left(\frac{\partial}{\partial\beta} - \vec{\tau} \cdot
\vec{\nabla}\right)\theta\nonumber$$
$$Q^\dagger = \frac{\partial}{\partial\theta}\left(\vec{\tau} \cdot
\vec{x} + \beta\right) \mp
\theta^\dagger\left(\frac{\partial}{\partial\beta} - \vec{\tau} \cdot
\vec{\nabla}\right)\eqno(68)$$
$$J^a = \sbody12 \left[\frac{\partial}{\partial\theta} \tau^a \theta +
\theta^\dagger \tau^a \frac{\partial}{\partial\theta^\dagger}\right] +
\left(-i \vec{x} \times \vec{\nabla}\right)^a\nonumber$$
$$Z = \pm \left(\theta^\dagger \frac{\partial}{\partial\theta^\dagger} -
\theta \frac{\partial}{\partial\theta} \right).\nonumber$$
Two quantities that commute with $Q$ are
$\vec{x}^{\,2} - \beta^2 \pm 
2\theta^\dagger \theta$ and $\displaystyle{\theta^\dagger
\frac{\partial}{\partial \theta^\dagger} + \theta
\frac{\partial}{\partial\theta} + \vec{x} \cdot \vec{\nabla} +
\beta\frac{\partial}{\partial\beta}}$. The changes induced by $Q$ and
$Q^\dagger$ on $x^a$ and $\theta$ are
$$\delta x^a = \left[ \xi^\dagger Q - Q^\dagger \xi , x^a \right] 
= 
\pm \left(\theta^\dagger \tau^a \xi -\xi^\dagger \tau^a \theta
\right)\nonumber$$
$$\delta \theta = \left[ \xi^\dagger Q - Q^\dagger\xi , \theta
\right]  =  \left(\vec{\tau} \cdot \vec{x} +
\beta\right)\xi .\eqno(69)$$

The physical interpretation of $\beta$ is not clear. Also, it is not
apparent how to construct a superfield containing component fields (such
as appear in (56)) that constitute an irreducible representation of the
supersymmetry transformation induced by (68). This may be due to an
inability to find a Grassmann operator that anticommutes with $Q$ and
$Q^\dagger$ as represented in (68).

\subsection{$AdS_2$}

In conjunction with the superalgebra of (19) we have a superspace
composed of Bosonic coordinates $x^a$ and Fermionic coordinates
$\theta_i$ where $\theta_i$ is a two component Majorana spinor. A suitable
representation of this superalgebra in superspace is provided by
$$J^{ab} = \frac{\partial}{\partial\theta} \Sigma^{ab}\theta - \left(x^a
\partial^b - x^b \partial^a\right)\nonumber$$
$$Q = \gamma^a \partial_a \theta + \gamma^a x_a
\frac{\partial}{\partial\tilde{\theta}}\eqno(70)$$
$$\tilde{Q} = -\tilde{\theta} \gamma^a \partial_a +
\frac{\partial}{\partial\theta} \gamma^a x_a .\nonumber$$
No analogue of the variable $\beta$ appearing in (68) is needed. Two
quantities that commute with $Q$ appearing in (70) are
$$R^2 = x^a x_a - \tilde{\theta}\theta\eqno(71)$$
and
$$\Delta = x^a\partial_a + \theta_i \frac{\partial}{\partial \theta_i} =
x^a\partial_a + \tilde{\theta}_i
\frac{\partial}{\partial\tilde{\theta}_i} .\eqno(72)$$
We also can define
$$D = -\gamma^a \partial_a \theta + \gamma^a x_a
\frac{\partial}{\partial\tilde{\theta}}\nonumber$$
$$\tilde{D} = \tilde{\theta} \gamma^a \partial_a +
\frac{\partial}{\partial\theta} \gamma^a x_a .\eqno(73)$$
Several useful relations are
$$\left[ D_j , \Delta \right] = 0\nonumber$$
$$\left\lbrace Q_i , \tilde{D}_j \right\rbrace = -
2\overline{\Delta}\delta_{ij}\nonumber$$
$$\left(\overline{\Delta} \equiv x^a \partial_a + \sbody32 \theta_i
\frac{\partial}{\partial\theta_i}\right)\eqno(74)$$
$$\left[ Q_i , \overline{\Delta} \right] = \sbody12 D_i\nonumber$$
$$\left[Q_i , D \tilde{D}\right] = 2\left[D_i , \overline{\Delta}
\right] = Q_i .\nonumber$$
From (72), we see that one can define a supersymmetric invariant
condition
$$\Delta\Phi = \omega \Phi\eqno(75)$$
on a scalar superfield $\Phi$ . This is analogous to the homogeneity
condition used in $dS_4$ space in [33].  A suitable invariant action can
be taken to be 
$$S_1 = \int d^3x d^2\theta \delta (R^2 - a^2) \Phi (\tilde{D} D +
\rho)\Phi .\eqno(76)$$
If now
$$\Phi = \phi + \tilde{\lambda} \theta + F\tilde{\theta}\theta
\eqno(77)$$
then by (75)
$$(x \cdot \partial - \omega)\phi = (x \cdot \partial + 1 - \omega )
\lambda = (x \cdot + 2 - \omega )F = 0 .\eqno(78)$$
Noting that
$$\tilde{D}D = -x^2 \frac{\partial}{\partial\theta}
\frac{\partial}{\partial\tilde{\theta}} + \frac{1}{2x^2} \left[ L^{ab}
L_{ab} + 2(x \cdot \partial )^2 + 2(x \cdot \partial
)\right]\tilde{\theta}\theta\nonumber$$
$$+ 2x \cdot \partial + 2 \tilde{\theta}\left(- x \cdot \partial +
\Sigma^{ab} L_{ab}\right)
\frac{\partial}{\partial\tilde{\theta}} - 3 \tilde{\theta} \frac{\partial}
{\partial\tilde{\theta}}\eqno(79)$$
and
$$\delta\left(R^2 - a^2\right) = \delta\left(x^2 - a^2\right) -
\tilde{\theta}\theta \delta^\prime \left(x^2 - a^2\right)\nonumber$$
we see that the component form of (76) is
$$S_1 = \int d^3x \left\lbrace \delta\left(x^2 - a^2\right)\left[-
\tilde{\lambda}\left(\Sigma^{ab} L_{ab} + \frac{\rho -
3}{2}\right)\lambda \right.\right.\eqno(80)$$
$$+ \frac{1}{2x^2} \phi \left(L^{ab} L_{ab} + 2\omega (1-\omega )\right)
\phi - 2x^2 F\nonumber$$
$$\left. + 2 (\rho - 1) \phi F\right] +\delta^\prime \left(x ^2 - a^2
\right) \left[2x^2 \phi F\right.\nonumber$$
$$\left. \left. - (\rho + 2\omega )\phi^2 \right]\right\rbrace
.\nonumber$$
Upon integrating over $\sqrt{x^2}$, we find that the action on the
$AdS_2$ surface is
$$S_1 = \int d^2A a^2 \left[- \tilde{\lambda} \left(\Sigma^{ab} L_{ab} +
\frac{\rho - 3}{2}\right)\lambda\right. \eqno(81)$$
$$+ \frac{1}{2a^3} \phi \left(L^{ab}L_{ab} + 2 \omega(1 + 5 \omega + \rho ) + \rho\right)
\phi\nonumber$$
$$\left. -2a^2F^2 +(2\rho - 3 -2\omega )\phi F\right] .\nonumber$$
If $\delta\Phi = \left[ \tilde{\xi} Q, \Phi \right]$, then it follows
that
$$\delta\phi = i\tilde{\xi}\gamma \cdot x \lambda\nonumber$$
$$\delta\tilde{\lambda} = i\tilde{\xi}\gamma \cdot (\partial\phi +
2xF)\eqno(82)$$
$$\delta F = -\sbody{i}{2} \tilde{\xi} \gamma\cdot
\partial\lambda .\nonumber$$
There is no immediate connection between the actions of (64) and (81),
although the changes of (82) can be identified with those of (65)
provided $i\gamma \cdot x \lambda \rightarrow \Psi$, $\omega \rightarrow
-1 - \chi$ and $2x^2 F\rightarrow - F$ in (82).

In place of (76) one could also consider the supersymmetric invariant
actions
$$S_2 = \int d^3x d^2\theta \delta \left(R^2 - a^2\right)\Phi(\tilde{Q}Q
+ \rho)\Phi\eqno(83a)$$
$$S_3 = \int d^3x d^2\theta \delta \left(R^2 -
a^2\right)\left[(\tilde{D}\Phi)(D\Phi) + \rho \Phi^2\right]\eqno(83b)$$
$$S_4 = \int d^3x d^2\theta \delta \left(R^2 -
a^2\right)\left[(\tilde{Q}\Phi)(Q\Phi) + \rho \Phi^2\right]\eqno(83c)$$
as well as supersymmetric invariant interactions
$$S_I = \lambda_N \int d^3x d^2\theta \delta\left(R^2 -
a^2\right)\Phi^N .\eqno(84)$$
(Actually, (83a) and (83c) are identical as $\left[Q_i , R^2\right] = 0$.)

Establishing supersymmetric invariance of $S_1 \cdots S_4$ in (76) and
(83) is not easily done if one works directly in terms of component
fields as in (81). However one can argue as follows to establish this
invariance. The expansion $(\tilde{D}D + \rho)\Phi$ is itself a scalar
superfield given by
$$P + \tilde{S}\theta + R\tilde{\theta}\theta = \left(-2x^2 F +
2\omega\phi\right) + \tilde{\theta}\left(2\Sigma^{ab}L_{ab} -
3\right)\lambda\nonumber$$
$$+ \tilde{\theta}\theta \left[ -2(1 + \omega) \phi + \frac{1}{2x^2}
\left(L^{ab}L_{ab} + 2\omega (1+\omega)\right)\phi\right]\eqno(85)$$
and hence the change in the product of two scalar fields is given by
$$\left[\tilde{\xi}Q, \left(P_1 + \tilde{\theta} S_1 + R_1
\tilde{\theta}\theta\right) \left( P_2 +\tilde{\theta} S_2 + R_2
\tilde{\theta}\theta \right)\right]\nonumber$$
$$= \left[\tilde{\xi} Q , P_1 P_2 + \left(P_1 \tilde{S}_2 + P_2
\tilde{S}_1 \right)\theta + \left(P_1 R_2 + P_2R_1
\right.\right.\nonumber$$
$$\left.\left. - \sbody12 \tilde{S}_1 S_2 \right)\tilde{\theta}\theta\right]
\nonumber$$
$$= -\sbody12 \partial_a \left[ \tilde{\xi}\gamma^a \left(P_1 S_2 + P_2
S_1\right)\right]\tilde{\theta}\theta + O(\theta) .\nonumber$$
Since this is the divergence of a current at order
$\tilde{\theta}\theta$, $S_1$, $S_2$, $S_3$ and $S_4$ are all
supersymmetric invariant actions.

\section{Discussion}

In this paper we have presented the simplest superalgebras associated
with spaces of constant curvature in two, three and four dimensions. In
this way, some of the superalgebras considered in [35] are exhibited
explicitly.  In addition, some novel superalgebras (eg, that of (13))
have been noted whcih do not fall into the categories considered by
Nahm [37] or Pilch et al. [38].

We have also considered the two dimensional models in more detail. In
particular, we have examined their representations. One peculiar result
occurs in the case of the superalgebra of eq. (7) associated with $S_2$, namely 
the requirement that states
have positive definite norm restricts the angular momentum to be less
than a value given in terms of 
the eigenvalues of the internal symmetry generator (cf eq. (38)). In
addition, component field models associated with superalgebras related
to $AdS_2$ and $S_2$ have been devised, and superfield models invariant
under the $AdS_2$ superalgebra are given.

Clearly more work remains to be done. Formulating models in three and
four dimensions is a high priority. Considering spaces of constant
curvature in dimensions higher than four also merits attention. It may
also be possible to relate the algebra of (13) to non-commutative
geometry [39].
These questions currently are under consideration.

\section{Appendix}
\subsection{Three Dimensions}

In $3 + 0$ dimensions, the Dirac matrices can be identified with the Pauli spin
matrices $\tau_{ij}^a$. These satisfy
$$\tau^a \tau^b = \delta^{ab} + i\epsilon^{abc}\tau^c\eqno(A.1)$$
$$\tau_{ij}^a \tau_{k\ell}^a = 2 \delta_{i\ell} \delta_{kj} -
\delta_{ij} \delta_{k\ell}\eqno(A.2)$$
$$\tau_{ij}^a \delta_{k\ell} + \tau_{k\ell}^a \delta_{ij} =
\tau_{i\ell}^a \delta_{kj} + \tau_{kj}^a \delta_{i\ell}\eqno(A.3)$$
$$\epsilon^{abc} \tau_{ij}^b \tau_{k\ell}^c = i \left(\tau_{i\ell}^a
\delta_{kj} - \tau_{kj}^a \delta_{i\ell} \right).\eqno(A.4)$$
Charge conjugation of a spinor $\psi$ is given by $\psi_c =
C(\psi^\dagger)^T$ where
$$C^{-1} \tau^\mu C = -(\tau^\mu)^T ;\eqno(A.5)$$
$C$ is taken to be
$$C = \tau^2 = C^\dagger = C^{-1} = -C^T = -C^* .\eqno(A.6)$$

In $2 + 1$ dimensions, we use the metric $\eta_{ab} = {\rm{diag}} (+, -, +)$ and
choose
$\gamma_1 = i\tau_1$, $\gamma_2 = \tau_2$, $\gamma_3 = i\tau_3$ so that
$$\gamma_a \gamma_b = -\eta_{ab} - i\epsilon_{abc} \gamma^c\eqno(A.7)$$
as $\epsilon^{123} = +1$. We take also
$$\Sigma_{ab} = \sbody14 \left[\gamma_a , \gamma_b\right] = -
\sbody{i}{2} \epsilon_{abc} \gamma^c .\eqno(A.8)$$
Since
$$\gamma_2\Sigma_{ab}\gamma_2 = -\Sigma_{ab}^T = -\Sigma_{ab}^\dagger
\eqno(A.9)$$
both $\overline{Q}Q$ and $\tilde{Q}Q$ are invariant under the
transformation
$$Q \rightarrow \exp \left( -\sbody12 \omega_{ab} \Sigma^{ab}
\right)Q\eqno(A.10)$$
where
$$\overline{Q} = Q^\dagger \gamma_2\;\;\;\;\; , \tilde{Q} = Q^T
\gamma_2 .\eqno(A.11a,b)$$
The matrix $C$ is defined by (A.5) and (A.6) in both $3dM$ and $3dE$; in
$3dM$ $\psi_C = C\overline{\psi}^T$.

If in $2 + 1$ dimensions, $\theta$, $\xi$ and $\chi$ are all Majorana (viz. $\theta =
\theta_c$, $\xi = \xi_c$, $\chi = \chi_c$) then we have
$$\left( \tilde{\xi} \chi \right) = \left( \tilde{\chi}\xi\right) = \left(\tilde{\xi}\chi
\right)^\dagger\eqno(A.12)$$
$$\left(\tilde{\xi}\gamma^a \chi\right) = -\left(\tilde{\chi}\gamma^a\xi\right) \eqno(A.13)$$
$$\left(\tilde{\xi} \theta\right)\left(\tilde{\theta}\chi\right) = -\sbody12 \left(\tilde{\xi}\chi
\right)\left(\tilde{\theta}\theta \right).\eqno(A.14)$$

In $3 + 0$ dimensions, $(\psi_c)_c = -\psi$ so one cannot impose the Majorana
condition; spinors can be Dirac, or 
alternatively, one can have a pair of ``symplectic Majorana'' spinors
$Q_i$ satisfying
$$Q_i = \epsilon_{ij}(Q_c)_j\eqno(A.15)$$
where $\epsilon_{ij} = -\epsilon_{ji}$, $\epsilon_{12} = 1$.

\subsection{Four Dimensions}

In $4 + 0$ dimensions, we choose the following representation for the Dirac matrices
$$\gamma^i = \left(
\begin{array}{cc}
0 & i\tau^i \\
-i\tau^i & 0\end{array}\right)\;\;\;\;
\gamma^0 = \left(
\begin{array}{cc}
0 & 1 \\
1 & 0\end{array}\right)\eqno(A.16)$$
$$\Sigma^{\mu\nu} = -\sbody14 \left[\gamma^\mu ,\gamma^\nu\right]\;\;\;
\;\;\;\;\gamma^5 = \gamma^1\gamma^2\gamma^3\gamma^4 = \left(\begin{array}{cc}
-1 & 0\\
0 & 1\end{array}\right)\nonumber$$
so that
$$\left\lbrace \gamma^\mu , \gamma^\nu \right\rbrace = 2g^{\mu\nu}
,\;\;\;
\left[\Sigma^{\alpha\beta} ,\gamma^\gamma\right] =
\delta^{\alpha\gamma}\gamma^\beta - \delta^{\beta\gamma}\gamma^\alpha
,\;\;\;
\left[\Sigma^{\mu\nu} ,\Sigma^{\lambda\sigma} \right] =
\delta^{\mu\lambda} \Sigma^{\nu\sigma} + \cdots \eqno(A.17)$$

Contracting the expansion
$$\Sigma_{ij}^{\mu\nu} \Sigma_{k\ell}^{\mu\nu} = a_{i\ell} \delta_{kj}
+ a_{i\ell}^5 \gamma_{kj}^5 + a_{i\ell}^\mu \gamma_{kj}^\mu + a_{i\ell}^{\mu 5}
\left(\gamma^\mu \gamma^5\right)_{kj}\nonumber$$
$$+ a_{i\ell}^{\mu\nu} \Sigma_{kj}^{\mu\nu}\eqno(A.18)$$
with $\delta_{kj}$, $\gamma_{kj}^5$, $\gamma^\mu_{kj}$, $(\gamma^\mu \gamma^5)_{kj}$
and $\Sigma_{kj}^{\mu\nu}$ in turn leads to 
$$\Sigma_{ij}^{\mu\nu} \Sigma_{k\ell}^{\mu\nu} = -\sbody12 \Sigma_{i\ell}^{\mu\nu}
\Sigma_{kj}^{\mu\nu} - \sbody34 \left(\delta_{i\ell} \delta_{kj} +
\gamma_{i\ell}^5 \gamma_{kj}^5 \right) \eqno(A.19)$$
so that
$$\Sigma_{ij}^{\mu\nu} \Sigma_{k\ell}^{\mu\nu} + \Sigma_{i\ell}^{\mu\nu}
\Sigma_{kj}^{\mu\nu} = -\sbody12\left[\left(\delta_{ij} \delta_{k\ell} +
\delta_{i\ell} \delta_{kj}\right) + \left(\gamma_{ij}^5 \gamma^5_{k\ell}
+ \gamma^5_{i\ell} \gamma^5_{kj}\right)\right].\eqno(A.20)$$
Similarly, one has
$$\delta_{ij}\delta_{k\ell} = \sbody14 \left[ \delta_{i\ell}\delta_{kj} + 
\gamma_{i\ell}^5 \gamma_{kj}^5 + \gamma_{i\ell}^\mu \gamma_{kj}^\mu - (\gamma^\mu
\gamma^5)_{i\ell} (\gamma^\mu \gamma^5)_{kj}\right.\nonumber$$
$$\left.-2 \Sigma_{i\ell}^{\mu\nu} \Sigma_{kj}^{\mu\nu}\right].\eqno(A.21)$$
Together, (A.20) and (A.21) give
$$2\left(\delta_{ij}\delta_{k\ell} + \delta_{i\ell}\delta_{kj}\right)
= \left(\gamma_{ij}^a \gamma_{k\ell}^a + \gamma_{i\ell}^a \gamma_{kj}^a\right)
 - \left[\left(\gamma^a
\gamma^5\right)_{ij} \left(\gamma^a \gamma^5\right)_{k\ell}\right.\nonumber$$
$$\left. +\left(\gamma^a\gamma^5\right)_{i\ell}
\left(\gamma^a\gamma^5\right)_{kj}\right] + 2 \left[\left(
\gamma^5\right)_{ij}
\left(\gamma^5\right)_{k\ell} + \left(\gamma^5\right)_{i\ell}
\left(\gamma^5\right)_{kj}\right].\eqno(A.22)$$
For charge conjugation, we now take
$$C \gamma^\mu C^{-1} = -(\gamma^\mu)^T\eqno(A.23)$$
and with $C = \gamma^0\gamma^2$,
$$C = -C^T = -C^\dagger = -C^{-1} = C^* .\eqno(A.24)$$
If $\psi_C = C(\psi^\dagger)^T$, then $(\psi_C)_C = -\psi$ so a spinor
cannot be Majorana; it can be Dirac, or we can have a pair of
symplectic Majorana spinors.

In $3 + 1$ dimensions, where $g^{\mu\nu} = {\rm{diag}}(+ - - -)$, we choose
$$\gamma^i = \left(\begin{array}{cc}
0 & \tau^i\\
-\tau^i & 0 \end{array}\right)\;\;\;\;
\gamma^0 = \left(\begin{array}{cc}
0 & 1\\
1 & 0\end{array}\right)\;\;\;\;
\gamma^5 = \left(\begin{array}
{cc}
-1 & 0\\
0 & 1\end{array}\right) .\eqno(A.25)$$
The charge conjugation matrix $C$ is taken to be $-i\gamma^0\gamma^2$ and so that if
$$\overline{\psi} = \psi^\dagger \gamma^0 \;\;\;\; , 
\psi_C = C(\overline{\psi})^T\eqno(A.25)$$
then it is apparent that the Majorana condition can be imposed. 

One can also
use 2-component notations for spinors in $3 + 1$ dimensions, using the conventions
of [35].  However, we find it convenient to avoid 
distinguishing between upper and lower case indices and to not use the
dot notation for indices; rather we choose to strictly employ matrix notation.

We first define
$$\sigma^\mu = (1, \vec{\tau})\;\;\;\; , \;\;\;\;\overline{\sigma}^\mu = (1, -\vec{\tau})
\nonumber$$
$$\sigma^{\mu\nu} = -\sbody14\left(\sigma^\mu\overline{\sigma}^\nu - \sigma^\nu
\overline{\sigma}^\mu\right), \;\;\;\;\; \overline{\sigma}^{\mu\nu} = -\sbody14\left(
\overline{\sigma}^\mu\sigma^\nu - \overline{\sigma}^\nu
\sigma^\mu\right).\nonumber$$
These satisfy the relations
$$\sigma^{\mu\dagger} = \sigma^\mu\;\; ,\; \;\;\overline{\sigma}^{\mu\dagger} = \overline{\sigma}^\mu\; ,
\;\;\;\sigma^{\mu\nu\dagger} = -\overline{\sigma}^{\mu\nu}\nonumber$$
$$\sigma_2 \sigma^\mu \sigma_2 = \overline{\sigma}^{\mu T}\;\;\;\;
\sigma_2\sigma^{\mu\nu}\sigma_2 = -\sigma^{\mu\nu T}\nonumber$$
$$\left[\sigma^{\mu\nu} , \sigma^{\lambda\sigma}\right] = g^{\mu\lambda} \sigma^{\nu\sigma}
- g^{\mu\sigma}\sigma^{\nu\lambda} + g^{\nu\sigma}\sigma^{\mu\lambda} -
g^{\nu\lambda}\sigma^{\mu\sigma}\nonumber$$
$$\sigma^{\mu\nu} \sigma^\lambda = \sbody12\left(g^{\mu\lambda} \sigma^\nu - g^{\nu\lambda}
\sigma^\mu + i\epsilon^{\mu\nu\lambda\rho}\sigma_\rho\right)\nonumber$$
$$\overline{\sigma}^\mu\sigma^{\nu\lambda} = -\sbody12\left(g^{\mu\nu}\overline{\sigma}^\lambda
- g^{\mu\lambda}\overline{\sigma}^\nu + i\epsilon^{\mu\nu\lambda\rho}
\overline{\sigma}_\rho \right) \;\;\; \left(\epsilon^{1230} = +1\right)\nonumber$$
$$\left(\sigma^{\mu\nu}\right)_{ij} \left(\sigma_{\mu\nu}\right)_{k\ell} = -2
\delta_{i\ell}\delta_{kj}+ \delta_{ij}\delta_{k\ell}\; ,\;\;\;
\left(\sigma^{\mu\nu}\right)_{ij} \left(\overline{\sigma}_{\mu\nu}\right)_{k\ell} = 0
\;\;\; 
\left(\sigma^{\mu}\right)_{ij} \left(\overline{\sigma}_{\mu}\right)_{k\ell} = 2
\delta_{i\ell}\delta_{kj} . \nonumber$$
Consider now 2-component spinors $\psi$ and $\chi$ and define
$$\tilde{\psi} = \psi^T\sigma_2\;\;\;\;
\tilde{\chi} = \chi^T \sigma_2\;\; .\nonumber$$
It follows from the above relations that if
$$\psi \rightarrow e^{\omega_{\mu\nu}\sigma^{\mu\nu}}\psi\;\;\;\;
\chi \rightarrow e^{\omega_{\mu\nu}\overline{\sigma}^{\mu\nu}}\chi\nonumber$$
then
$$\tilde{\psi} \rightarrow \tilde{\psi}e^{-\omega_{\mu\nu}\sigma^{\mu\nu}}\;\;\;
\tilde{\chi} \rightarrow \tilde{\chi}e^{-\omega_{\mu\nu}\overline{\sigma}^{\mu\nu}}\nonumber$$
$$\psi \rightarrow \psi^\dagger e^{-\omega_{\mu\nu}\overline{\sigma}^{\mu\nu}}
\;\;\;
\chi^\dagger \rightarrow \chi^\dagger e^{-\omega_{\mu\nu}\sigma^{\mu\nu}}\nonumber$$
$$\left(\sigma_2 \psi^{\dagger T}\right) \rightarrow e^{\omega_{\mu\nu}\overline{\sigma}^{\mu\nu}}
\left(\sigma_2\psi^{\dagger T}\right)\;\;\;
\left(\sigma_2 \chi^{\dagger T}\right) \rightarrow e^{\omega_{\mu\nu}\sigma^{\mu\nu}}
\left(\sigma_2\chi^{\dagger T}\right).\nonumber$$
It is apparent then that the following structures are Lorentz invariant:
$$\tilde{\psi}\psi\,,\;\;\;\; 
\tilde{\chi}\chi\, ,\;\;\;\; \psi^\dagger \overline{\sigma}_\mu \psi\, ,
\;\;\;\;\chi^\dagger \sigma_\mu \chi \, ,
\;\;\;\; \tilde{\psi}\sigma_{\mu\nu} \psi\, ,\;\;\;\; \tilde{\chi}\overline{\sigma}
_{\mu\nu} \chi .\nonumber$$
A suitable candidate for a pair of relativisticly invariant wave equations
are
$$\overline{\sigma}^\mu p_\mu \psi + m \chi + 0\nonumber$$
$$\sigma^\mu p_\mu \chi - m\psi = 0\nonumber$$
Together, these equations imply that
$$\left(p^2 - m^2\right)\psi = 0 = \left(p^2 - m^2\right)\chi\nonumber$$

For $2 + 2$ dimensions we employ
$$\gamma^1 = \left(\begin{array}{cc}
0 & 1\\
1 & 0\end{array}\right) \;\;\;
\gamma^2 = \left(\begin{array}{cc}
0 & i\tau^2\\
-i\tau^2 & 0\end{array}\right) \;\;\;
\gamma^3 = \left(\begin{array}{cc}
0 & \sigma^1\\
-\sigma^1 & 0\end{array}\right) \;\;\;
\gamma^4 = \left(\begin{array}{cc}
0 & \sigma^3\\
-\sigma^3 & 0\end{array}\right) \eqno(A.26)$$
with $g^{\mu\nu} = \rm{diag} (+, +, -, -)$. If now
$$C = -A = -i\gamma^3\gamma^4\eqno(A.27)$$
and $\psi_C = C(\overline{\psi})^T = C(\psi^\dagger A)^T = \psi^*$, then
it is
apparent that a spinor can now be both Majorana $(\psi = \psi_C)$ and
Weyl ($\psi = \pm \gamma^5\psi$ where $\gamma^5 = -
\gamma^1\gamma^2\gamma^3\gamma^4$).
The 2-component notation we use for $2 + 2$ dimensions is defined in [5].

\subsection{Five Dimensions}

In $5 + 0$ dimensions, we use the matrices of (A.16); the analogues of (A.17)
continue to hold. The following equations
are useful
$$\gamma^A\gamma^B\gamma^C = \delta^{AB}\delta^C - \delta^{AC}\gamma^B +
\delta^{BC}\gamma^A + \epsilon^{ABCDE}\Sigma^{DE}\eqno(A.28)$$
$$\gamma^A\gamma^B\gamma^C\gamma^D = \delta^{AB}\delta^{CD} -
\delta^{AC}\delta^{BD} + \delta^{AD}\delta^{BC}\nonumber$$
$$+\epsilon^{ABCDE} \gamma^E -
2\left[\delta^{AB}\Sigma^{CD}\right.\nonumber$$
$$+\delta^{AC}\Sigma^{DB} + \delta^{BC}\Sigma^{AD} +
\delta^{AD}\Sigma^{BC}\nonumber$$
$$\left. +\delta^{BD}\Sigma^{CA} +
\delta^{CD}\Sigma^{AB}\right]\eqno(A.29)$$
$$(\Sigma \cdot A)(\Sigma \cdot B) = -\sbody12 A \cdot B + \sbody14
\epsilon^{ABCDE} A^{AB} B^{CD} \gamma^E+ 2 A^{AC} B^{BC} \Sigma^{AB}
.\eqno(A.30)$$
Just as (A.18) can be used to derive (A.19), we can show
$$\Sigma_{ij}^{AB}\Sigma_{k\ell}^{AB} = -\sbody12 \Sigma_{i\ell}^{AB}
\Sigma_{kj}^{AB} - \sbody14 \gamma_{i\ell}^A \gamma_{kj}^A - \sbody54
\delta_{i\ell}\delta_{kj}\eqno(A.31)$$
$$\gamma_{ij}^A \gamma_{k\ell}^A = -\sbody34 \gamma_{i\ell}^A \gamma_{kj}^A
+ \sbody54 \delta_{i\ell}\delta_{kj} - \sbody12 \Sigma_{i\ell}^{AB}
\Sigma_{kj}^{AB} \eqno(A.32)$$
$$\delta_{ij}\delta_{k\ell} =\sbody14 \delta_{i\ell}\delta_{kj}
+ \sbody14 \gamma_{i\ell}^A \gamma_{kj}^A - \sbody12 \Sigma_{i\ell}^{AB}
\Sigma_{kj}^{AB} \eqno(A.33)$$
Together, (A.31) to (A.33) can be used to demonstrate that
$$\Sigma_{ij}^{AB} \Sigma_{k\ell}^{AB} + \Sigma_{i\ell}^{AB} \Sigma_{kj}^{AB}
= -\left(\delta_{ij} \delta_{k\ell} + \delta_{i\ell}\delta_{kj}\right)
= -\left(\gamma_{ij}^A \gamma_{k\ell}^A + \gamma_{i\ell}^A \gamma_{kj}^A\right).\eqno(A.34)$$

In $5 + 0$ dimensions we have
$$\gamma^0 = \left(\begin{array}{cc}
0 & 1\\
1 & 0\end{array}\right)\;\;\;\;
\gamma^i = \left(\begin{array}{cc}
0 & i\tau^i\\
-i\tau^i & 0\end{array}\right)\;\;\;\;
\gamma^5 = \left(\begin{array}{cc}
 -1 & 0\\
0 & 1\end{array}\right) \eqno(A.35)$$
and hence it is appropriate to define $\overline{\psi} = \psi^\dagger$ and to have
$C = -\gamma^1\gamma^3$ so that
$$\gamma^A = C^{-1} \gamma^{AT} C .\eqno(A.36)$$
We hence take $\psi_C = C\overline{\psi}^T$; $\psi$ cannot be Majorana, 
but may be Dirac or symplectic Majorana.
In $4 + 1$ dimensions we employ
$$\gamma^0 = \left(\begin{array}{cc}
0 & 1\\
1 & 0\end{array}\right)\;\;\;\;
\gamma^i = \left(\begin{array}{cc}
0 & \tau^i\\
-\tau^i & 0\end{array}\right)\;\;\;\;
\gamma^5 = \left(\begin{array}{cc}
 -i & 0\\
0 & i\end{array}\right) \eqno(A.37)$$
define $\overline{\psi} = \psi^\dagger \gamma^0$ and $\psi_C = C\overline{\psi}^T$ with
$$C = -i\gamma^0 \gamma^2 \gamma^5 = C^\dagger = C^{-1} = -
C^T\eqno(A.38)$$
so that $\gamma^A = C^{-1}(\gamma^A)^T C$. Since $(\psi_C)_C = -\psi$,
spinors can be either Dirac or symplectic Majorana, but not Majorana.

In $3 + 2$ dimensions, we have
$$\gamma^0 = \left(\begin{array}{cc}
0 & 1\\
1 & 0\end{array}\right)\;\;\;\;
\gamma^i = \left(\begin{array}{cc}
0 & \sigma^i\\
-\sigma^i & 0\end{array}\right)\;\;\;\;
\gamma^5 = \left(\begin{array}{cc}
 -1 & 0\\
0 & 1\end{array}\right) \eqno(A.39)$$
so that it is appropriate that we take
$\overline{\psi} = \psi^{\dagger}\left(\gamma^1\gamma^2\gamma^3\right)$
and $\psi_C = C\overline{\psi}^T$ where
$$C = i\gamma^2\gamma^3\eqno(A.40)$$
so that $\gamma^A = C^{-1}(\Gamma^A)^T C$ as in $4 + 1$ dimension. The Majorana
condition $\psi = \psi_C$ is consistent as $(\psi_C)_C = \psi$ in $3 + 2$ dimensions.

An additional identity that is useful in five dimensions is
$$\left(\Sigma^{AB}C\right)_{ij} \left(\Sigma_{AB}\right)_{k\ell} +
\left(\Sigma^{AB}C\right)_{jk}\left(\Sigma_{AB}\right)_{i\ell} +
\left(\Sigma^{AB}C\right)_{ki}\left(\Sigma_{AB}\right)_{j\ell} =
0;\eqno(A.41)$$
this holds in $5 + 0$ dimensions, $4 + 1$ dimensions and $(3 + 2)$ dimensions.  It is proven by
expanding the left side of (A.41) in the form $P_{ij}\delta_{k\ell} +
Q_{ij}^A(\gamma_A)_{k\ell} + R_{ij}^{AB}(\Sigma_{AB})_{k\ell}$;
contraction with $\delta_{\ell k}$, $(\gamma_C)_{\ell k}$ and
$(\Sigma_{CD})_{\ell k}$ leads to $P = Q^A = R^{AB} = 0$.
\section{Acknowledgement}
We would like to thank NSERC for financial support. R. and D. MacKenzie had a 
number of helpful suggestions.

\end{document}